\documentstyle[preprint,prb,aps]{revtex}

\begin{document}

\draft

\title{Hydrodynamics of Spatially Ordered Superfluids}

\author{H.T.C. Stoof,$^1$ Kieran Mullen,$^2$ Mats Wallin,$^3$
        and S.M. Girvin$^4$}
\address{$^1$University of Utrecht, Institute for Theoretical
             Physics, Princetonplein 5, \\ P.O. Box 80.006,
             3508 TA  Utrecht, The Netherlands \\
         $^2$University of Oklahoma, Department of Physics,
             Norman, OK 73019-0225 \\
         $^3$Royal Institute of Technology, Department of
             Theoretical Physics, S-100 44 Stockholm, Sweden \\
         $^4$Indiana University, Department of Physics,
             Bloomington, IN 47405}

\maketitle

\begin{abstract}
We derive the hydrodynamic equations for the supersolid and superhexatic phases
of a neutral two-dimensional Bose fluid. We find, assuming that the normal part
of the fluid is clamped to an underlying substrate, that both phases can
sustain third-sound modes and that in the supersolid phase there are additional
modes due to the superfluid motion of point defects (vacancies and
interstitials).
\end{abstract}


\section{INTRODUCTION}
Liquid helium ($^4$He) has a reputation for being the first substance in which
one is able to observe many macroscopic quantum phenomena. In particular, it
was the first system that could sustain superfluid flow, \cite{bulk} and as a
consequence display a number of amazing properties such as second sound,
quantized vortices and the fountain effect. Furthermore, thin superfluid helium
films were the first two-dimensional systems experimentally proven to undergo a
Kosterlitz-Thouless
transition to the normal state. \cite{film} More recently it may
have been observed \cite{Mochel} that on weakly-binding substrates these films
are the first-known spatially ordered superfluids. \cite{VAV}

More precisely, measurements of the third-sound resonance
frequency (which is proportional to the square root of the superfluid density)
of submonolayer helium films on hydrogen and deuterium substrates apparently
indicate two independent Kosterlitz-Thouless transitions: the usual superfluid
to normal transition at a temperature $T_{KT}$ that obeys the expected
universal jump relation, \cite{NK} and a second new transition at a temperature
$T_c$ which is roughly $0.5\,\,T_{KT}$ for all coverages. The second transition
appears as a sharp (but not discontinuous) rise or dip in the superfluid
density depending on the substrate.

In an attempt to explain these experimental results we have
recently proposed that below the second critical temperature the superfluid
helium film is in a spatially ordered phase exhibiting both off-diagonal
(superfluid) and diagonal (hexatic) long-range order in the one-particle
density matrix. \cite{PRL} The main idea behind this proposal is that the
hexatic to fluid transition is known to be a Kosterlitz-Thouless transition
driven by disclination unbinding. \cite{NH}  (Disclinations are defects in the
orientational order of a crystal created by the insertion or removal of a wedge
of atoms, as shown in Fig.\ 1.) Therefore, our physical picture of the
experiments is that at sufficiently low temperatures the film is in a
superhexatic phase with only a dilute gas of bound vortices and bound
disclinations present due to thermal fluctuations. For entropic reasons the
disclinations then unbind at $T_c$, leading to a transition from a superhexatic
to a superfluid phase since the vortices remain bound at this transition and
the presence of free disclinations destroys the hexatic long-range order. At
$T_{KT}$ the vortices then also unbind and the film is finally forced into the
normal liquid phase.

Of course, to make sure that the above picture is qualitatively correct we must
also consider the interaction between vortices and disclinations. This is even
more pressing if one realizes that in a supersolid phase (where all
disclination pairs are themselves bound into pairs or triples) this interaction
is of long range and depends logarithmically on the distance between the two
kinds of defects. Fortunately, it turns out that this is no longer true in the
superhexatic phase due to the screening of the interaction by the surrounding
gas of disclination pairs. A renormalization-group analysis actually shows that
the vortex-disclination interaction is irrelevant and that the two separate
Kosterlitz-Thouless transitions indeed survive. Nevertheless, the superfluid
density is influenced in a non-universal way by the unbinding of the
disclinations and Monte-Carlo simulations even show that on the basis of our
hypothesis a rough qualitative agreement with the experiments of Chen and
Mochel can be obtained. \cite{PRL}

However, to definitely identify the phase below $T_c$ more detailed information
is needed. As a first step towards this goal we here present the
two-dimensional hydrodynamic equations of a superhexatic by describing the
superhexatic as a supersolid with free dislocations (i.e.\ disclination pairs
\cite{NH}). As a result of this approach we will also be able to consider the
hydrodynamics of the supersolid phase, for which there is at present a renewed
interest both in the context of Josephson-junction arrays \cite{Anne} and solid
$^4$He. \cite{LG} Moreover, spatially ordered superfluid states have recently
been proposed to be also relevant for the fractional quantum Hall effect,
\cite{B} since this effect can be understood as a condensation of composite
bosons. \cite{Z} We therefore believe that the methods developed below might,
if extended to bosons interacting with a Chern-Simons gauge field, also be used
to obtain a description of the dynamics of such exotic quantum Hall states.

We have organized the paper in the following manner. In Sec.\ \ref{PD} we first
consider the normal solid and hexatic phases by formulating a gauge theory that
describes the phonons, the dislocations and the interaction between them. From
this theory we then deduce in Sec.\ \ref{SOP} for both phases the dynamics of
the appropriate hydrodynamic degrees of freedom. In Sec.\ \ref{SP} we
incorporate the effects of the additional superfluid order parameter
\cite{fluid} into the hydrodynamic equations derived in Sec.\ \ref{SOP} and
discuss the various long-wavelength modes in the supersolid and superhexatic
phases obtained in this manner. We conclude in Sec.\ \ref{DC} with a discussion
on the possible relevance of our work to future experiments on submonolayer
helium films and with a physical interpretation of our results.

\section{GAUGE THEORY OF PHONONS AND DISLOCATIONS}
\label{PD}
In this section we will derive the long-wavelength (quantum) dynamics of the
solid and hexatic phases. The discussion closely follows work by Kleinert,
\cite{K1} save that we will not include higher gradient elasticity. This leads
to a considerable simplification of the theory but implies that we cannot
properly treat the dynamics of the disclinations. Fortunately, for our purposes
only the dynamics of the dislocations is of importance and this simplification
is justified.

\subsection{Solid}
\label{S}
In the case of an isotropic crystal, the action for the
displacement field $u_i(\vec{x},\tau)$ in the presence of a pair of
dislocations is given by \cite{K1}
\begin{equation}
\label{action}
S[u_i] = \int_0^{\hbar \beta} d\tau \int d\vec{x}~
\left\{ \frac{\rho}{2} (\partial_{\tau}u_i - \beta_i)^2
           + \mu \left(u_{ij} - \frac{\beta_{ij} + \beta_{ji}}{2}
                 \right)^2
           + \frac{\lambda}{2} (u_{ii} - \beta_{ii})^2 \right\}~,
\end{equation}
where $u_{ij} = (\partial_i u_j + \partial_j u_i)/2$ is the strain tensor,
$\mu$ and $\lambda$ are the usual Lam\'e coefficients \cite{La} and $\rho$ is
the average mass density. The unphysical (and singular) contributions arising
from the multivaluedness of $u_i(\vec{x},\tau)$ are compensated by the
quantities $\beta_j$ and $\beta_{ij}$ (also known as the `plastic distortion').
Their relationship to the defects is best explained by the Volterra
construction. \cite{K2} Let $\cal C$ be a small loop bounding a section of two
dimensional crystal that is excised from the whole (cf. Fig.\ 2). The edges of
the loop are drawn together and form a line $\cal L$. This line may be time
dependent, and its definition is not unique. However, the topological defects
(i.e.\ two dislocations with opposite Burgers' vectors) associated with the
distortion of the surface are always located at the endpoints of $\cal L$. If
$\pm \vec{B}$ are the Burgers' vectors of the dislocations constituting the
pair and if $\vec{v}$ is their velocity then
$\beta_{ij} = \delta_i({\cal L})B_j$ and
$\beta_j = - v_i \delta_i({\cal L})B_j$. The delta function $\delta_i({\cal
L})$ is singular on the time-dependent Volterra cutting line ${\cal L}$ of the
dislocations and is directed along the normal vector. If the cutting line
${\cal L}$ is parameterized by $\vec{x}(s,\tau)$ with $0 \leq s \leq 1$, this
means mathematically that
\begin{equation}
\delta_i({\cal L}) = - \epsilon_{ij} \int_0^1 ds~
    \frac{\partial x_j(s,\tau)}{\partial s}~
       \delta(\vec{x} - \vec{x}(s,\tau))~,
\end{equation}
where $\epsilon_{ij}$ is the two-dimensional antisymmetric tensor. Note that
the dislocations are assumed to be able to move freely, without any friction,
through the crystal because the equations of motion for the displacement field
allow for time-dependent solutions that precisely correspond to such evolutions
of the crystal. \cite{Na} We will come back to the issue of friction in Sec.\
\ref{SOP} when we consider the effects of dissipation.

We now first perform a Hubbard-Stratonovich transformation by introducing the
auxillary variable $\vec p$ (representing the momentum density) and adding the
quadratic term
\begin{eqnarray}
\int_0^{\hbar \beta} d\tau \int d\vec{x}~
  \frac{1}{2\rho} (p_i - i\rho(\partial_{\tau}u_i - \beta_i))^2
                                                 \nonumber
\end{eqnarray}
to the action, which may now be rewritten as
\begin{eqnarray}
S[p_i,u_i] = \int_0^{\hbar \beta} d\tau \int d\vec{x}~
\left\{ \frac{p_i^2}{2\rho} \right.
          &+& \mu \left(u_{ij} - \frac{\beta_{ij} +
                                       \beta_{ji}}{2} \right)^2
                                                 \nonumber \\
          &+& \left. \frac{\lambda}{2} (u_{ii} - \beta_{ii})^2
           - ip_i(\partial_{\tau}u_i - \beta_i) \right\}~.
\end{eqnarray}
Integrating out $\vec{p}$ would return the original action up to an unimportant
constant. In a similar manner we then also introduce the symmetric stress
tensor $\sigma_{ij}$, to decouple the terms quadratic in the strain. This
results in
\begin{eqnarray}
S[p_i,\sigma_{ij},u_i] =
              \int_0^{\hbar \beta} d\tau \int d\vec{x}~
\left\{ \frac{p_i^2}{2\rho} \right.
          &+& \frac{1}{4\mu} \left(\sigma_{ij}^2 -
                \frac{\nu}{1+\nu} \sigma_{ii}^2 \right)
                                                \nonumber \\
&-& \left. ip_i(\partial_{\tau}u_i - \beta_i)
         + i\sigma_{ij} \left( u_{ij} -
                        \frac{\beta_{ij} + \beta_{ji}}{2}
                                            \right) \right\}~,
\end{eqnarray}
with $\nu = \lambda/(2\mu + \lambda)$. The partition function is now given by
the functional integral
\begin{equation}
Z = \int d[p_i] \int d[\sigma_{ij}] \int d[u_i]~
  \exp \left\{-\frac{1}{\hbar} S[p_i,\sigma_{ij},u_i] \right\}~,
\end{equation}
where the integration over $\sigma_{ij}$ is only over the symmetrical part
since we have not included higher gradient elasticity.

We can now perform the integration over the displacement field. Because the
action is linear in $u_i$ this simply leads to the constraint
\begin{equation}
\label{con}
\partial_{\tau}p_j = \partial_i \sigma_{ij}~.
\end{equation}
This constraint can be automatically satisfied if we introduce the vector field
$A_j$ and the tensor field $A_{ij}$ by setting
$\sigma_{ij} = \epsilon_{ik} \partial_k A_j
             + \epsilon_{ki} \partial_\tau A_{kj}$
and $p_j = \epsilon_{ki} \partial_i A_{kj}$. Substituting these relations into
the action we find that the interaction between the gauge fields (i.e.\ the
phonons) and the dislocations is
given by
\begin{equation}
\label{Sint}
S_{int}[A_{ij},A_j] = \int_0^{\hbar \beta} d\tau \int d\vec{x}~
   \{ -iA_i \alpha_i + iA_{ij} J_{ij} \}~,
\end{equation}
where after several partial integrations the unphysical singularities of
$\beta_i$ and $\beta_{ij}$ have disappeared and only the dislocation density
and the dislocation current density
remain. Introducing also the function
$\delta({\cal P}) = \delta(\vec{x}(1,\tau)) -
                                 \delta(\vec{x}(0,\tau))$,
which denotes the difference between a delta function at one end of the cutting
line ${\cal L}$ and a delta function at the other end, these densities and
currents can conveniently be written as
$\alpha_j = \delta({\cal P})B_j$ and
$J_{ij} = - v_i \delta({\cal P})B_j$, respectively. As a direct consequence of
the above definitions they obey the conservation law
\begin{equation}
\label{Claw}
\partial_{\tau} \alpha_j = \partial_i J_{ij}~.
\end{equation}

In addition, the dynamics of the phonons is determined by the  remaining
quadratic terms in the action which expressed in terms of the gauge fields
$A_j$ and $A_{ij}$ yield
\begin{equation}
S_0[A_{ij},A_j] = \int_0^{\hbar \beta} d\tau \int d\vec{x}~
  \left\{ \frac{(\epsilon_{ik} \partial_k A_{ij})^2}{2\rho}
          + \frac{1}{4\mu} \left(\sigma_{ij}^2 -
               \frac{\nu}{1+\nu} \sigma_{ii}^2 \right) \right\}~,
\end{equation}
with $\sigma_{ij}$ equal to
$\epsilon_{ik}(\partial_k A_j - \partial_\tau  A_{kj})$.
Comparing this result with Eq.\ (\ref{action}) we observe that the stress and
the physical part of the strain
$u^{Phys}_{ij} \equiv u_{ij} - (\beta_{ij} + \beta_{ji})/2$
are related by
$\sigma_{ij} = 2\mu u^{Phys}_{ij} + \lambda \delta_{ij}
                                                u^{Phys}_{kk}$
and therefore by
\begin{equation}
\label{uphys}
u^{Phys}_{ij} = \frac{1}{2\mu} \left( \sigma_{ij}
            - \frac{\nu}{1+\nu} \delta_{ij} \sigma_{kk} \right)~.
\end{equation}
We will have need of the latter relation in Sec.\ \ref{SOP}, when we discuss
hydrodynamics. A more formal way to justify it is to add to the action
$S[u_i]$ a source term
\begin{eqnarray}
\int_0^{\hbar \beta} d\tau \int d\vec{x}~
  K_{ij} \left(u_{ij} - \frac{\beta_{ij} + \beta_{ji}}{2} \right)
    = \int_0^{\hbar \beta} d\tau \int d\vec{x}~
                              K_{ij} u^{Phys}_{ij}   \nonumber
\end{eqnarray}
and perform the same manipulations as before. We then find that the source
$K_{ij}$ indeed couples linearly to the right-hand side of Eq.\ (\ref{uphys}).

Following Kleinert, we now notice that the above theory has
a gauge symmetry as a result of the fact that the gauge fields
$A_i$ and $A_{ij}$ are not uniquely determined if the stresses $\sigma_{ij}$
and momenta $p_i$ are known. Indeed,
$\sigma_{ij}$ and $p_i$ are invariant under the gauge transformation $A_i
\rightarrow A_i + \partial_{\tau} \Lambda_i$ and $A_{ij} \rightarrow A_{ij} +
\partial_i \Lambda_j$. Hence $S_0[A_{ij},A_j]$ is also invariant. Moreover, due
to the conservation law in Eq.\ (\ref{Claw}), the interaction
$S_{int}[A_{ij},A_j]$ is invariant too.

To calculate the partition function we therefore need some gauge-fixing
procedure. The symmetry of $\sigma_{ij}$ requires that
\begin{equation}
\epsilon_{ij} \sigma_{ij} = \partial_j A_j
                               - \partial_{\tau}(A_{jj}) = 0~.
\end{equation}
We would now like to write the gauge fields as the appropriate derivatives of
unconstrained fields. Using the above gauge symmetry we can always take
$A_i = \epsilon_{ij} \partial_j \chi$ and $A_{ii}=0$.
This, however, does not completely fix the gauge because these conditions are
still invariant under the smaller group of transformations
$\chi \rightarrow \chi + \partial_{\tau} \Lambda$ and
$A_{ij} \rightarrow A_{ij} + \partial_i (\epsilon_{jk}
                             \partial_k \Lambda)$.
To see more clearly the consequences of this residual symmetry we
expand $A_{ij}$ into its longitudinal and transverse components (with respect
to both indices), i.e.
\begin{equation}
A_{ij} = \partial_i (\partial_j A^{LL})
         + \partial_i (\epsilon_{jk} \partial_k A^{LT})
         + \epsilon_{ik} \partial_k (\partial_j A^{TL})
         + \epsilon_{ik} \partial_k
                             (\epsilon_{jl} \partial_l A^{TT})~,
\end{equation}
where we have introduced four new fields. The tracelessness of $A_{ij}$ can
then be fulfilled by taking $A^{LL} = -A^{TT}$. In addition, the residual gauge
symmetry can now be written as
$\chi \rightarrow \chi + \partial_{\tau} \Lambda$ and
$A^{LT} \rightarrow A^{LT} + \Lambda$. This shows that instead of the fields
$\chi$ and $A^{LT}$ we must use the gauge-invariant field $\chi' \equiv \chi
- \partial_{\tau} A^{LT}$ together with $\Lambda$ as integration variables. The
associated change of
measure can be incorporated in the normalization and the same is true for the
`volume' $\int d[\Lambda]$ of the residual gauge group because the action is
gauge invariant and therefore cannot depend on $\Lambda$. After this
gauge-fixing procedure the partition function thus becomes
\begin{equation}
Z = \int d[A^{TT}] \int d[A^{TL}] \int d[\chi']~
  \exp \left\{-\frac{1}{\hbar} \left(S_0[A^{TT},A^{TL},\chi'] +
                 S_{int}[A^{TT},A^{TL},\chi'] \right) \right\}~.
\end{equation}
Note that we are left with three physical degrees of freedom, which is the
correct number in two dimensions since $\sigma_{ij}$ and $p_i$ contain in
principle a total of five degrees of freedom but we have two constraints in
Eq.\ (\ref{con}). Note also that the transformation from $\sigma_{ij}$ and
$p_i$ to $A^{TT}$, $A^{TL}$ and $\chi'$ is a linear one so that the Jacobian
involved in the calculation of the partition function is simply an unimportant
constant. In particular, the stress is given by
\begin{equation}
\label{stress}
\sigma_{ij}= \epsilon_{ik}\epsilon_{j\ell}\partial_k \partial_\ell \chi'
  + \partial_\tau\left(
    \partial_i \partial_j A^{TL}
  + \epsilon_{ik} \partial_k \partial_j A^{TT}
  + \epsilon_{jk} \partial_k \partial_i A^{TT}
        \right)~,
\end{equation}
which is manifestly symmetric in $i$ and $j$.

A straightforward calculation now shows that the free part of the action is
\begin{eqnarray}
\label{s0}
S_0[A^{TT},A^{TL},\chi'] &=&
   \int_0^{\hbar \beta} d\tau \int d\vec{x}~
     \left\{ \frac{1}{2\mu} (\partial_{\tau} \partial^2 A^{TT})^2
           + \frac{1}{2\rho} (\partial_i \partial^2 A^{TT})^2
     \right. \nonumber \\
 &+& \left.
     \frac{1}{4\mu(1+\nu)} (\partial_{\tau} \partial^2 A^{TL})^2
           + \frac{1}{2\rho} (\partial_i \partial^2 A^{TL})^2
           + \frac{1}{4\mu(1+\nu)} (\partial^2 \chi')^2
     \right. \nonumber \\
 &-& \left.
     \frac{1}{2\mu} \frac{\nu}{1+\nu}
        (\partial_{\tau} \partial^2 A^{TL})(\partial^2 \chi')
     \right\}~.
\end{eqnarray}
It contains four modes: The part involving $A^{TT}$ has a pair of modes
(corresponding to $\pm \vec{k}$) with
$\omega^2 = \mu \vec{k}^2/\rho$. These modes therefore represent the transverse
phonons with a speed of sound $\sqrt{\mu/\rho}$. The part involving $A^{TL}$
and $\chi'$ has another pair of modes with a dispersion obeying
$\omega^2 = (2\mu + \lambda) \vec{k}^2/\rho$.
These represent the longitudinal phonons with a speed of sound $\sqrt{(2\mu +
\lambda)/\rho}$. Interestingly, these results can be understood much more
easily if we introduce the field
\begin{equation}
\label{chidp}
\chi'' \equiv \chi' - \nu \partial_{\tau} A^{TL}~,
\end{equation}
since then the above action becomes
\begin{eqnarray}
\label{uncoupled}
S_0[A^{TT},A^{TL},\chi''] &=&
   \int_0^{\hbar \beta} d\tau \int d\vec{x}~
     \left\{ \frac{1}{2\mu} (\partial_{\tau} \partial^2 A^{TT})^2
           + \frac{1}{2\rho} (\partial_i \partial^2 A^{TT})^2
     \right. \nonumber \\
 &+& \left.
     \frac{1-\nu}{4\mu} (\partial_{\tau} \partial^2 A^{TL})^2
           + \frac{1}{2\rho} (\partial_i \partial^2 A^{TL})^2
           + \frac{1}{4\mu(1+\nu)} (\partial^2 \chi'')^2
     \right\}~,
\end{eqnarray}
so that the fields are completely uncoupled. Notice that the $\chi''$ field has
no kinetic term, which explains why we obtained above only four modes instead
of six, as might have been expected in first instance.

Furthermore, if we introduce the usual Burgers' field $\vec{b}(\vec{x},\tau)$
for the total dislocation density, which is nothing more than the sum of the
density $\alpha_i$ over all dislocation pairs, then the interaction with the
dislocations acquires the form
\begin{equation}
S_{int}[A^{TT},A^{TL},\chi'] =
  \int_0^{\hbar \beta} d\tau \int d\vec{x}~
    i \left\{ \chi' \epsilon_{ij} \partial_j b_i
              - A^{TT} \partial_{\tau} \partial_i b_i
      \right\}~,
\end{equation}
where we have made use of Eq.\ (\ref{Claw}) to express the longitudinal part
of the current density $J_{ij}$ in terms of the time derivative of $b_i$.
Decomposing $\vec{b}$ into its transverse and longitudinal parts, i.e.\
$b_i = \partial_i b^L + \epsilon_{ij} \partial_j b^T$, the interaction finally
becomes
\begin{equation}
\label{int}
S_{int}[A^{TT},A^{TL},\chi'] =
  \int_0^{\hbar \beta} d\tau \int d\vec{x}~
    i \left\{ \chi' \partial^2 b^T
              - A^{TT} \partial_{\tau} \partial^2 b^L
      \right\}~.
\end{equation}
The total action $S = S_0 + S_{int}$ reduces for time-independent
configurations to the one we previously used for a discussion of the critical
properties of the superhexatic. \cite{PRL} Integrating out the fields $A^{TT}$
and $\chi'$ we can now find the time-dependent interaction among the
dislocations. Physically, these interactions are thus associated with phonon
exchange and the time dependence arises due to the finite speeds of sound. This
picture also explains why the effective action for $\chi'$ contains just
one pair of modes: The self-interaction of the
transverse dislocation density can only be mediated by
longitudinal phonons.

\subsection{Hexatic}
\label{Hex}
Up to this point the dislocation density has not been an independent
dynamical variable, since we have specified the positions of the dislocations
at all times and thus neglected the influence of the phonon dynamics on
their motion. However, to describe the hexatic phase we want to
integrate out the dislocations in the
plasma (or continuous) approximation. \cite{NH} For that we need the free
action of the field $\vec{b}$. Here we can again make use of the results
obtained by Kleinert, who showed that the energy associated with the nonlinear
stresses at the heart of the defect can be lumped into a `core contribution' to
the action. \cite{K1,K3} In our notation this contribution becomes
\begin{eqnarray}
\label{dislo}
S_0[b_i] &=& \int_0^{\hbar \beta} d\tau \int d\vec{x}~
    \frac{E_c}{2} b_i
     \left( \frac{\rho}{\mu}
            \frac{\partial_{\tau}^2}{\partial^2} + 1 \right) b_i
                                                     \nonumber \\
 &=& \int_0^{\hbar \beta} d\tau \int d\vec{x}~
    \frac{E_c}{2} \left\{
       b^T \left( \frac{\rho}{\mu} \partial_{\tau}^2
                                 + \partial^2 \right) b^T
     + b^L \left( \frac{\rho}{\mu} \partial_{\tau}^2
                                 + \partial^2 \right) b^L
                   \right\} ~.
\end{eqnarray}
This action represents free propagation of the dislocation density fluctuations
which, as mentioned previously, are permitted by the classical equations of
motion \cite{Na} and neglects dissipative coupling of the dislocation cores to
the phonons.

Integrating out the Burgers' field using Eqs.\ (\ref{int}) and (\ref{dislo}),
we obtain the following results. The effective action for $A^{TT}$ becomes
\begin{eqnarray}
S^{eff}[A^{TT}] =
  \int_0^{\hbar \beta} d\tau \int d\vec{x}~
    \left\{ \frac{1}{2\mu} (\partial_{\tau} \partial^2 A^{TT})^2
    \right. + \frac{1}{2\rho} (\partial_i \partial^2 A^{TT})^2
                                 \hspace*{1.0in} \nonumber \\
   + \left.
     \frac{1}{2E_c} (\partial_{\tau} \partial_i A^{TT})
       \left(
         \frac{\rho}{\mu}\frac{\partial_{\tau}^2}{\partial^2} + 1
       \right)^{-1}
                    (\partial_{\tau} \partial_i A^{TT})
    \right\}~.
\end{eqnarray}
As shown in Fig.\ 3a, it contains two pairs of modes which for
$\vec{k}^2 \gg \mu/2E_c$ all have a dispersion obeying
$\omega^2 \simeq \mu \vec{k}^2/\rho$. However, for
$\vec{k}^2 \ll \mu/2E_c$, one pair of modes has a dispersion
$\omega^2 \simeq \mu^2/E_c\rho + 2\mu\vec{k}^2/\rho$ with a gap whereas
the other pair of modes is gapless with
$\omega^2 \simeq 2E_c \vec{k}^4/\rho$. This is consistent with our expectation
that in the hexatic phase there should only be one pair of transverse
gapless modes with a softer dispersion than that of the transverse phonon
modes in a true solid.

The effective action for $\chi'$ and $A^{TL}$ in the hexatic phase is
\begin{eqnarray}
S^{eff}[A^{TL},\chi'] &=&
   \int_0^{\hbar \beta} d\tau \int d\vec{x}~
     \left\{
     \frac{1}{4\mu(1+\nu)} (\partial_{\tau} \partial^2 A^{TL})^2
     \right.
          + \frac{1}{2\rho} (\partial_i \partial^2 A^{TL})^2
                                                   \nonumber \\
         &+& \frac{1}{4\mu(1+\nu)} (\partial^2 \chi')^2
          - \frac{1}{2\mu} \frac{\nu}{1+\nu}
            (\partial_{\tau} \partial^2 A^{TL})(\partial^2 \chi')
                                                   \nonumber \\
  &+& \left.
      \frac{1}{2E_c} (\partial_i \chi')
       \left(
         \frac{\rho}{\mu}\frac{\partial_{\tau}^2}{\partial^2} + 1
       \right)^{-1} (\partial_i \chi')
       \right\}~.
\end{eqnarray}
Integrating out also $A^{TL}$ we finally arrive at the effective action for
$\chi'$. It reads
\begin{eqnarray}
S^{eff}[\chi'] = \int_0^{\hbar \beta} d\tau \int d\vec{x}~
     \left\{ \frac{1}{4\mu(1+\nu)} (\partial^2 \chi')^2
       + \frac{1}{2E_c} (\partial_i \chi')
           \left(
             \frac{\rho}{\mu}\frac{\partial_{\tau}^2}{\partial^2}
                                                              + 1
           \right)^{-1} (\partial_i \chi')
     \right.       \hspace*{0.5in}        \nonumber \\
     \left.
       + \frac{1}{2} \left(\frac{1}{2\mu}
                              \frac{\nu}{1+\nu} \right)^2
          (\partial_{\tau} \partial^2 \chi')
            \left( \frac{\partial_{\tau}^2}{2\mu(1+\nu)}
                      + \frac{\partial^2}{\rho} \right)^{-1}
             (\partial_{\tau} \partial^2 \chi')
     \right\}
\end{eqnarray}
and also contains two pairs of modes (cf. Fig.\ 3b).
For $\vec{k}^2 \gg \mu/2E_c$ we
recover of course the ordinary sound dispersions
$\omega^2 \simeq 2\mu\vec{k^2}/(\rho(1-\nu))
                            = (2\mu +\lambda)\vec{k}^2/\rho$
and $\omega^2 \simeq \mu\vec{k}^2/\rho$. However, in the limit
$\vec{k}^2 \ll \mu/2E_c$ these evolve into a pair of gapped modes with
$\omega^2 \simeq 2\mu^2/(E_c\rho(1-\nu))$ and a pair of propagating modes
with $\omega^2 \simeq 2\mu(1+\nu)\vec{k}^2/\rho$, respectively.
Clearly, the same mode structure is
also present in the effective action for $A^{TL}$ (obtained by integrating out
$\chi'$ instead of $A^{TL}$) which indicates that in the
hexatic phase the longitudinal velocity is renormalized downwards to
$\sqrt{2\mu(1+\nu)/\rho}$.

\section{HYDRODYNAMICS OF SPATIALLY ORDERED PHASES}
\label{SOP}
We now turn to the linear hydrodynamics of the solid and hexatic phases that
follows from the theory presented above. For the sake of clarity, and because
it will turn out to be less important for our purposes, we will not discuss
temperature fluctuations in the following. However, having derived the
relevant energy
densities in Secs. \ref{S} and \ref{Hex} it is in principle straightforward to
include temperature fluctuations in our theory and, in particular, to arrive at
the extension of the hydrodynamic equations presented below that is required if
one wants to consider also the hydrodynamic mode due to energy conservation.
After the equations of motion for the hydrodynamic variables are determined,
we can find the propagating and diffusive modes. This is done as before,
by Fourier transforming the equations of motion and determining the dispersion
$\omega(k)$.
Propagating modes appear as complex roots of a characteristic equation and
will always occur in pairs. Each physically distinct propagating excitation
such as longitudinal or transverse sound corresponds therefore to two
roots or modes.

We start by considering the mass-density fluctuation $\delta\rho$ above the
average mass density $\rho$ and initially neglect the possible presence of
vacancies and interstitials. To lowest order in the strain, the density
fluctuation equals $-\rho u^{Phys}_{ii}$ so up to that order we obtain
\begin{equation}
\partial_{\tau} \delta\rho = - \rho \partial_{\tau} u^{Phys}_{ii}
  = - \frac{\rho}{2(\mu + \lambda)} \partial_{\tau}
                                                  \sigma_{ii}~,
\end{equation}
if we make use of Eq.\ (\ref{uphys}) to relate the stress and the strain. Using
also the decomposition
$\sigma_{ii} = \partial^2 \chi' + \partial_{\tau}
                                             \partial^2 A^{TL}$ from
Eq.\ (\ref{stress})
we may write this as a pair of continuity equations
\begin{mathletters}
\label{hydro}
\begin{equation}
\partial_{\tau} \delta\rho = \partial_i g_i^L~,
\end{equation}
\begin{equation}
\partial_{\tau} g_j^L = \partial_i \pi_{ij}^D~,
\end{equation}
\end{mathletters}

\noindent
where the longitudinal momentum density is given by
\begin{equation}
g_i^L = - \frac{\rho}{2(\mu + \lambda)} \partial_{\tau}
           \partial_i (\chi' + \partial_{\tau} A^{TL})
\end{equation}
and the diagonal part of the stress tensor by
\begin{equation}
\pi_{ij}^D = - \frac{\rho \delta_{ij}}{2(\mu + \lambda)}
             \partial_{\tau}^2 (\chi' + \partial_{\tau} A^{TL})~.
\end{equation}

In the absence of defects, the hydrodynamic quantities $g_i^L$ and
$\pi_{ij}^D$ are precisely equal to the longitudinal part of $p_i$ and the
diagonal part of $\sigma_{ij}$, respectively. This can be seen in the
following manner. In an ideal solid we have no defects, and variation
of the action in Eq.\ (\ref{uncoupled}) gives $\chi''=0$ or
$\chi'=\nu \partial_\tau A^{TL}$, which we may use to eliminate $\chi'$.
The equation of motion for $A^{TL}$ generated in this way is
\begin{equation}
\label{ATLmotion}
\partial_{\tau}^2 A^{TL} = - \frac{2\mu + \lambda}{\rho}
                                          \partial^2 A^{TL}.
\end{equation}
If we substitute this back into the definitions of $g_i^L$ and $\pi_{ij}^D$
we obtain the longitudinal part of $p_i$ and the
diagonal part of $\sigma_{ij}$ as given in section \ref{PD}.
In the presence of defects with their own dynamics this is no longer true,
since the dislocation density couples to the gauge fields and alters
the equations of motion. To avoid confusion about this point we have,
therefore, introduced a new notation for the hydrodynamic
momentum density and stress tensor which we will use for the rest of
the paper.

We also note that the above equations
are not Galilean invariant and are therefore only valid in a specific reference
frame. This is a result of the fact that the gauge theory of Sec.\ \ref{PD} has
implicitly used the existence of an ideal lattice with respect to which the
displacements $\vec{u}(\vec{x},\tau)$ are defined. \cite{K2} Hence, the
prefered reference frame corresponds to that frame in which this ideal lattice
is at rest. This is the case for all the hydrodynamic equations presented
below.

In this ideal solid without interstitials or vacancies the pressure
fluctuation (following from
$\pi_{ij}^D = - \delta_{ij} \delta p$) equals
\begin{equation}
\label{press}
\delta p = \frac{\rho}{2(\mu + \lambda)}
              (1 + \nu) \partial_{\tau}^3 A^{TL}
         = \frac{\rho}{2\mu + \lambda} \partial_{\tau}^3
                                                         A^{TL}
\end{equation}
and the mass-density fluctuation becomes
\begin{equation}
\delta\rho = - \frac{\rho}{2(\mu + \lambda)}
              (1 + \nu) \partial_{\tau} \partial^2 A^{TL}
            = - \frac{\rho}{2\mu + \lambda}
                        \partial_{\tau} \partial^2 A^{TL}~.
\end{equation}
Substituting the equation of motion Eq.\ (\ref{ATLmotion})
for $A^{TL}$ into Eq.\ (\ref{press}), we obtain the desired constitutive
equation
\begin{equation}
\delta p = \frac{2\mu + \lambda}{\rho} \delta\rho~.
\end{equation}
Together with the hydrodynamic equations (\ref{hydro}) this correctly leads to
the sound equation
\begin{equation}
\label{sound}
\partial_{\tau}^2 \delta\rho
    = - \frac{2\mu + \lambda}{\rho} \partial^2 \delta\rho
    = - c_{||}^2 \partial^2 \delta\rho~,
\end{equation}
with $c_{||}$ the longitudinal sound velocity.

However, as stressed first by Martin, Parodi, and Pershan \cite{M} and again by
Zippelius, Halperin, and Nelson \cite{ZHN} we are not in general allowed to
assume that the crystal is ideal, without vacancies or interstitials. We must
include the effects of (long-wavelength) fluctuations in the net defect density
$n_{\Delta}$, which is defined as the density
of vacancies minus the density of interstitials. To do so we can make use of
the fact that in a hexagonal system these defects can be regarded as a `bound
state' of three dislocations with radial Burgers' vectors pointing
symmetrically outward (interstitial) or inward (vacancy). \cite{N} This is
illustrated for an interstitial in Fig.\ 4. As a result the interaction of the
net defect density with the gauge fields is given by
\begin{equation}
S_{int}[A^{TT},A^{TL},\chi'] =
          \int_0^{\hbar \beta} d\tau \int d\vec{x}~
      i \frac{\gamma_{\Delta}}{\mu} n_{\Delta} \partial^2
                                                          \chi'~,
\end{equation}
where $V_0$ denotes the area deficit induced by a defect in an otherwise
perfect crystal and $\gamma_\Delta =\mu {c_{||}^2} {V_0}/{2}{c_{\perp}^2}$.
We can verify this result by noting that in the static case (and $n_{\Delta}
\rightarrow in_{\Delta}$ because of our conventions in the imaginary time
formalism of Sec.\ \ref{PD}) the Euler-Lagrange equation for the Airy stress
function, following from the action in Eq.\ (\ref{s0}) together with the above
interaction, becomes
$\partial^2 \chi = 2(\mu + \lambda) V_0 n_{\Delta}$
which correctly leads to
$\int d\vec{x}~u^{Phys}_{ii} = V_0 \int d\vec{x}~n_{\Delta}$.
Furthermore, the free action of the defects becomes
(cf. Eq.\ (\ref{dislo}))
\begin{equation}
\label{free}
S_0[n_{\Delta}] = \int_0^{\hbar \beta} d\tau \int d\vec{x}~
    \frac{E_{\Delta}}{2} n_{\Delta}
     \left( \frac{\rho}{\mu} \frac{\partial_{\tau}^2}{\partial^2}
             + 1 \right) n_{\Delta}~,
\end{equation}
with $E_{\Delta}$ of order $E_c V_0$.

Redoing our calculations with $n_{\Delta}$ non-zero, we find that
$n_{\Delta}$ displaces the $\chi'$ field. Therefore $\chi''$ in Eq.\
(\ref{chidp}) is also non-zero. Moreover, we now obtain instead of Eq.\
(\ref{sound}) the coupled set of equations
\begin{mathletters}
\label{sounds}
\begin{equation}
\partial_{\tau}^2 \delta\rho =
   - c_{||}^2 \left(
                1 + \frac{\nu \gamma_{\Delta}^2}
                                         {E_{\Delta} \mu}
              \right) \partial^2 \delta\rho
   + i \gamma_{\Delta}
              \left(
                1 - \frac{2 \gamma_{\Delta}^2}
                                         {E_{\Delta} \mu}
              \right) \partial^2 n_{\Delta}~,
\end{equation}
\begin{equation}
\partial_{\tau}^2 n_{\Delta} =
   - c_{\perp}^2 \left(
                   1 + \frac{2 \gamma_{\Delta}^2}
                                         {E_{\Delta} \mu}
                 \right) \partial^2 n_{\Delta}
   + i \frac{\gamma_{\Delta}\lambda}
                   {E_{\Delta}\rho^2} \partial^2 \delta\rho~,
\end{equation}
\end{mathletters}

\noindent
for the longitudinal degrees of freedom. Note that the density fluctuation
$\delta\rho$ receives a contribution from both the lattice vibrations as well
as
from the net defect density, since
\begin{equation}
\delta\rho =
   - \frac{1}{c_{||}^2} \partial_{\tau} \partial^2 A^{TL}
   + \frac{2i \gamma_{\Delta}}{c_{||}^2} n_{\Delta}~.
\end{equation}
As a result the longitudinal momentum density has also two contributions
\begin{equation}
g^L_i =
   - \frac{1}{c_{||}^2} \partial_i \partial_{\tau}^2 A^{TL}
   - \frac{2i \gamma_{\Delta}}{c_{||}^2} J^L_i~,
\end{equation}
where $\vec{J}^L$ is the longitudinal part of the net defect current density
obeying the continuity equation
$\partial_{\tau} n_{\Delta} = - \partial_i J^L_i$.

This almost completes our discussion of the hydrodynamical description (without
dissipation) of the solid phase. However, we have not yet obtained the
transverse modes. From our expressions for the strain tensor $u_{ij}$ one can
easily show that in the solid phase the transverse part of the displacement
field is given by
\begin{equation}
\label{disp}
u^T_i = \frac{1}{\mu} \epsilon_{ij}
                           \partial_j (\partial_{\tau} A^{TT})
      = \frac{1}{\rho c_{\perp}^2} \epsilon_{ij}
                           \partial_j (\partial_{\tau} A^{TT})~,
\end{equation}
where $c_{\perp}$ is the transverse speed of sound. Hence, the transverse
dynamics of the lattice is solely determined by the transverse phonons and we
have the additional hydrodynamic equation
\begin{equation}
\partial_{\tau}^2 A^{TT} =
                    - c_{\perp}^2 \partial^2 A^{TT}~,
\end{equation}
which is completely uncoupled from the previous ones and in particular does not
depend on the net defect density $n_{\Delta}$. Moreover, if we introduce the
standard hexatic order parameter $\vartheta_6$, which is equal to the local
bond angle and may therefore be written as
\begin{equation}
\vartheta_6 \equiv \frac{1}{2} \epsilon_{ij} \partial_i u_j
            = - \frac{1}{2\rho c_{\perp}^2}
                     (\partial_{\tau} \partial^2 A^{TT})~,
\end{equation}
the equation for $A^{TT}$ is equivalent to
\begin{equation}
\partial_{\tau}^2 \vartheta_6 =
                    - c_{\perp}^2 \partial^2 \vartheta_6~,
\end{equation}
so that $\vartheta_6$ can also be used to describe the transverse phonons.

{}From Eq.\ (\ref{disp}) we also find that the transverse part of the momentum
density is given by
\begin{equation}
g_i^T = \frac{1}{c_{\perp}^2} \epsilon_{ij}
                   \partial_j (\partial_{\tau}^2 A^{TT})~.
\end{equation}
Therefore the stress tensor has the nondiagonal contribution
\begin{equation}
\pi_{ij}^{ND} = - \frac{1}{c_{\perp}^2} \epsilon_{ij}
                                  (\partial_{\tau}^3 A^{TT})
              = 2\rho c_{\perp}^2 \epsilon_{ij} \vartheta_6~,
\end{equation}
and both the longitudinal as well as the transverse hydrodynamic equations in
the solid phase can be summarized by
\begin{mathletters}
\begin{equation}
\frac{\partial \delta\rho}{\partial t} = -\nabla \cdot \vec{g}~,
\end{equation}
\begin{equation}
\label{momS}
\frac{\partial \vec{g}}{\partial t} =
      - c^2 \nabla \delta\rho
      - \gamma_{\Delta} \nabla n_{\Delta}
      + 2\rho c_{\perp}^2 \nabla \times \vartheta_6~,
\end{equation}
\begin{equation}
\label{transS}
\frac{\partial \vartheta_6}{\partial t} =
                   \frac{1}{2\rho} \nabla \times \vec{g}~,
\end{equation}
\begin{equation}
\frac{\partial n_{\Delta}}{\partial t} =
      -\nabla \cdot \vec{J}~,
\end{equation}
\begin{equation}
\label{longS}
\frac{\partial \vec{J}}{\partial t} =
      - c_{\Delta}^2 \nabla n_{\Delta}
      + \gamma \nabla \delta\rho~,
\end{equation}
\end{mathletters}

\noindent
after a transformation to real time, which in this case not only means that
$\tau \rightarrow it$ but also
$\vec{g} \rightarrow i\vec{g}$ and
$n_{\Delta} \rightarrow in_{\Delta}$. Moreover, note that the constants
$c$, $\gamma_{\Delta}$, $c_{\Delta}$, and $\gamma$ should here be
interpreted as renormalized quantities which are determined in terms of the
microscopic parameters of our gauge theory by a comparison with
Eq.\ (\ref{sounds}).

Now we are ready to discuss dissipation. In principle dissipation has already
been included because there is a coupling between the net defect density
$n_{\Delta}$ and the phonons. Hence if for example an interstitial were, in a
discrete picture, to tunnel from one location to another there would be a
`shake up' of the phonon field. However, if we treat $n_{\Delta}$ as a smooth
continuously varying field, the action in Eq.\ (\ref{free}) is quadratic and
the bilinear coupling
$n_{\Delta} \partial^2 \chi'$ produces only mixing of the collective modes
but no real dissipation. Therefore, we choose to include effective dissipation
in the same manner as explained in detail by Zippelius, Halperin and Nelson.
\cite{ZHN} Using their notation we first of all add to the right-hand side of
Eq.\ (\ref{momS}) the terms
$(\eta \nabla^2 \vec{g}
   + \zeta \nabla (\nabla \cdot \vec{g}) )/\rho$
associated with the dissipative part of the stress tensor $\pi_{ij}$ and
representing viscous diffusion of the momentum density.

Next the question arises how we need to modify
Eq.\ (\ref{longS}). This equation is a result of the fact that we have allowed
the dislocations, and hence the interstitials and vacancies, to move freely
through the lattice and used Eq.\ (\ref{free}) for the free action of the
defects. If the defects effectively experience friction (for example due to
their interaction with the phonons), then it is more appropriate to add a
Leggett friction term \cite{TL} and use
\begin{equation}
\label{diss}
S_0[n_{\Delta}] = \int_0^{\hbar \beta} d\tau \int d\vec{x}~
    \frac{E_{\Delta}}{2} n_{\Delta}
     \left( \frac{\rho}{\mu}
       \frac{\partial_{\tau}^2}{\partial^2}
         + i \frac{\rho}{\mu} \xi \partial_{\tau} + 1
     \right) n_{\Delta}
\end{equation}
instead. The dispersions then indeed obey
$\omega^{\pm} \simeq \pm c_{\perp} k - i \xi k^2/2$ at long wavelengths, and we
must add the term
$\xi \nabla (\nabla \cdot \vec{J})$ to the right-hand side of Eq.\
(\ref{longS}). If we further assume that the transverse part of the defect
current density behaves as in a gas and simply diffuses to zero with a
diffusion constant $\kappa$, we obtain in total
\begin{mathletters}
\label{HS}
\begin{equation}
\frac{\partial \delta\rho}{\partial t} = -\nabla \cdot \vec{g}~,
\end{equation}
\begin{equation}
\frac{\partial \vec{g}}{\partial t} =
      - \frac{B}{\rho} \nabla \delta\rho
      - \gamma_{\Delta} \nabla n_{\Delta}
      + 2\rho c_{\perp}^2 \nabla \times \vartheta_6
      + \frac{\eta}{\rho} \nabla^2 \vec{g}
      + \frac{\zeta}{\rho} \nabla (\nabla \cdot \vec{g})~,
\end{equation}
\begin{equation}
\frac{\partial \vartheta_6}{\partial t} =
                  \frac{1}{2\rho} \nabla \times \vec{g}~,
\end{equation}
\begin{equation}
\frac{\partial n_{\Delta}}{\partial t} =
      -\nabla \cdot \vec{J}~,
\end{equation}
\begin{equation}
\label{diff}
\frac{\partial \vec{J}}{\partial t} =
      - c_{\Delta}^2 \nabla n_{\Delta}
      + \gamma \nabla \delta\rho
      + \kappa \nabla^2 \vec{J}
      + \xi \nabla (\nabla \cdot \vec{J})~,
\end{equation}
\end{mathletters}

\noindent
with $B = \rho \partial p/\partial\rho|_{n_{\Delta},T} =
                                          \rho c^2$
the appropriate isothermal bulk modulus in view of the fact that the pressure
is a function of both the particle density as well as the net defect density.
{}From thermodynamics we therefore also conclude that
$\gamma_{\Delta} = \partial p/\partial n_{\Delta}|_{\rho,T}$.

It is interesting to point out that these hydrodynamic equations differ from
the results obtained by Zippelius, Halperin, and Nelson. In particular, their
Eq.\ (3.32) differs from our
Eq.\ (\ref{diff}) and reads
\begin{equation}
\vec{J} =
   - \Gamma_{\Delta} \nabla
       \left( \frac{n_{\Delta}}{\chi_{\Delta}} -
                                  \gamma_{\Delta} \delta\rho
       \right)~.
\end{equation}
The difference can easily be traced back to the fact that Zippelius, Halperin,
and Nelson assume on phenomenological grounds that the dynamics of the net
defect density is purely diffusive. Indeed, we exactly reproduce their results
if we use in our calculation a free action of the form
\begin{equation}
S_0[n_{\Delta}] = \int_0^{\hbar \beta} d\tau \int d\vec{x}~
    \frac{E_{\Delta}}{2} n_{\Delta}
          \left(
            \frac{i\partial_{\tau}}{D_{\Delta} \partial^2} + 1
          \right) n_{\Delta}
\end{equation}
instead of Eq.\ (\ref{free}). We can therefore consider the hydrodynamic
equations of Zippelius, Halperin and Nelson as the overdamped (or classical)
limit of our Eq.\ (\ref{HS}). Clearly, Kleinert's more microscopic approach
does not lead to purely diffusive but in first instance to propagating behavior
of
the defects, which is appropriate for the quantum crystals of interest in Sec.\
\ref{SP}. We now turn to the modification of the above results in the hexatic
phase.

In the hexatic phase there are free dislocations present and $\chi''$ replaces
$n_{\Delta}$ as the appropriate dynamical degree of freedom. To see most
clearly how this comes about we will work perturbatively in $1/E_c$. Up to
first order in $1/E_c$
the effective action for $\chi''$ is
\begin{eqnarray}
\int_0^{\hbar \beta} d\tau \int d\vec{x}~
     \left\{ \frac{1}{4\mu(1+\nu)} (\partial^2 \chi'')^2
       + \frac{1}{2E_c} (\partial_i \chi'')
           \left(
             \frac{\rho}{\mu}
             \frac{\partial_{\tau}^2}{\partial^2} + 1
           \right)^{-1} (\partial_i \chi'')
          \right\}~,                             \nonumber
\end{eqnarray}
which upon Fourier transformation displays two modes with
$\omega^2 = c_{\perp}^2 \vec{k}^2 + 2\mu^2(1+\nu)/(E_c\rho)$. So in the
limit $E_c \rightarrow \infty$ (which physically means that we are looking at
the nonhydrodynamic regime
$\vec{k}^2 \gg \mu/2E_c$) we approximately have
\begin{equation}
\partial_{\tau}^2 \chi'' = - c_{\perp}^2 \partial^2 \chi''~,
\end{equation}
whereas the equations of motion for $A^{TL}$ and $A^{TT}$ are
\begin{equation}
\partial_{\tau}^2 A^{TL} =
                    - c_{||}^2 \nabla^2 A^{TL}
                    + \frac{\nu}{1-\nu^2} \partial_{\tau} \chi''
\end{equation}
and
\begin{equation}
\partial_{\tau}^2 A^{TT} =
                    - c_{\perp}^2 \nabla^2 A^{TT}~,
\end{equation}
respectively. For the mass-density fluctuation we now find
\begin{equation}
\delta\rho = - \frac{1}{c_{||}^2}
            \partial_{\tau} \partial^2 A^{TL}
          - \frac{\rho}{2(\mu + \lambda)} \partial^2 \chi''
\end{equation}
and for the stress tensor
\begin{equation}
\pi_{ij}^D = - \delta_{ij}
   \left\{
      \frac{1}{c_{||}^2} \partial_{\tau}^3 A^{TL}
    + \frac{\rho}{2(\mu + \lambda)} \partial_{\tau}^2 \chi''
   \right\}
         = - \delta_{ij}
   \left\{ c_{||}^2 \delta\rho
           + \frac{1}{2(1+\nu)} \partial^2 \chi''
   \right\}~.
\end{equation}

Putting all this together we obtain in first instance the following set of
hydrodynamic equations for the hexatic phase
\begin{mathletters}
\begin{equation}
\partial_{\tau} \delta\rho = \nabla \cdot \vec{g}^L~,
\end{equation}
\begin{equation}
\partial_{\tau} \vec{g}^L = - \nabla \delta p
    = - c_{||}^2 \nabla \delta\rho
      - \frac{1}{2(1+\nu)} \nabla (\nabla^2 \chi'')~,
\end{equation}
\begin{equation}
\partial_{\tau}^2 \chi'' = - c_{\perp}^2 \nabla^2 \chi''~,
\end{equation}
\begin{equation}
\partial_{\tau}^2 A^{TT} =
                    - c_{\perp}^2 \nabla^2 A^{TT}~.
\end{equation}
\end{mathletters}

\noindent
Combining the longitudinal and transverse parts as before, this equals
\begin{mathletters}
\begin{equation}
\partial_{\tau} \delta\rho = \nabla \cdot \vec{g}~,
\end{equation}
\begin{equation}
\label{mom}
\partial_{\tau} \vec{g} =
      - c_{||}^2 \nabla \delta\rho
      - \frac{1}{2(1+\nu)} \nabla (\nabla^2 \chi'')
      + 2\rho c_{\perp}^2 \nabla \times \vartheta_6~,
\end{equation}
\begin{equation}
\partial_{\tau}^2 \chi'' = - c_{\perp}^2 \nabla^2 \chi''~,
\end{equation}
\begin{equation}
\partial_{\tau} \vartheta_6 =
                   - \frac{1}{2\rho} \nabla \times \vec{g}~,
\end{equation}
\end{mathletters}

\noindent
and clearly reduces to the hydrodynamic equations for the ideal crystal if we
put $\chi''=0$.

We now have to consider how the above picture changes for a finite value of
$E_c$. Here we can use the results of
Sec.\ \ref{Hex}. In the hydrodynamic regime
$\vec{k}^2 \ll \mu/2E_c$ we saw that the transverse speed of sound was
renormalized to zero, because we found the quadratic (particle-like) dispersion
$\omega^2 = 2E_c \vec{k}^4/\rho$. As a result we have for the transverse
part of the hydrodynamic equations
\begin{equation}
\partial_{\tau}^2 A^{TT} =
                    \frac{2E_c}{\rho} \partial^4 A^{TT}~,
\end{equation}
which implies that in the right-hand side of Eq.\ (\ref{mom}) we must replace
$2\rho c_{\perp}^2 \nabla \times \vartheta_6$ by
$- 4 E_c \vec{e}_z \times \nabla (\nabla^2 \vartheta_6)$. This gives
\begin{mathletters}
\begin{equation}
\partial_{\tau} \vec{g}^T =
      - 4 E_c \vec{e}_z \times \nabla (\nabla^2 \vartheta_6)~,
\end{equation}
\begin{equation}
\partial_{\tau} \vartheta_6 =
                   - \frac{1}{2\rho} \nabla \times \vec{g}^T~,
\end{equation}
\end{mathletters}

\noindent
which is in complete agreement with Zippelius, Halperin, and Nelson if we
identify the Frank constant $K_A$ with $8E_c$.

For the longitudinal part we need to analyze the dynamics of $\chi''$ and
$A^{TL}$. A straightforward calculation shows that the effective action for
these fields contains precisely the four modes already found in Sec.\
\ref{Hex}. The propagating modes with $\omega^2 = 2\mu(1+\nu)
\vec{k}^2/\rho$ obey
$\chi' = \chi'' + \nu \partial_{\tau} A^{TL} =0$ and are therefore indeed
associated with density fluctuations proportional to $\partial^2 \chi''$. We
thus need to use a renormalized longitudinal speed of sound equal to
\begin{equation}
c = \sqrt{\frac{2\mu (1+\nu)}{\rho}}
       = \sqrt{\frac{2\mu}{\rho}
                  \frac{2\mu + 2\lambda}{2\mu + \lambda}}
\end{equation}
that is always smaller than the longitudinal speed of sound in the solid phase.
In fact, this actually exhausts the longitudinal hydrodynamic modes since the
other modes in the effective action for $\chi''$ and $A^{TL}$ are gapped. As a
result we now obtain in real time the following set of hydrodynamic equations
for the hexatic phase
\begin{mathletters}
\begin{equation}
\frac{\partial \delta\rho}{\partial t}
                                = - \nabla \cdot \vec{g}~,
\end{equation}
\begin{equation}
\label{momH}
\frac{\partial \vec{g}}{\partial t} =
      - c^2 \nabla \delta\rho
      - \frac{K_A}{2}
           \,\,\vec{e}_z \times \nabla (\nabla^2 \vartheta_6)~,
\end{equation}
\begin{equation}
\label{transH}
\frac{\partial \vartheta_6}{\partial t} =
                    \frac{1}{2\rho} \nabla \times \vec{g}~,
\end{equation}
\end{mathletters}

\noindent
not including dissipation.

To include dissipation we again follow Zippelius, Halperin and Nelson and add
to the right-hand side of Eq.\ (\ref{momH}) the terms $(\eta \nabla^2 \vec{g}
        + \zeta \nabla (\nabla \cdot \vec{g}) )/\rho$.
However, we do not add the term $\kappa \nabla^2 \vartheta_6$ to the right-hand
side of Eq.\ (\ref{transH}) because, just as in the solid phase, the
dissipation of the transverse modes is already accounted for in the term $\eta
\nabla^2 \vec{g}$ that is added to the momentum equation. Put differently, a
term of the form $\kappa \nabla^2 \vartheta_6$ can be absorbed by an
appropriate redefinition of $K_A$ and $\eta$. Again introducing the isothermal
bulk modulus
$B = \rho dp/d\rho|_T = \rho c^2$ we then find
\begin{mathletters}
\label{HH}
\begin{equation}
\frac{\partial \delta\rho}{\partial t}
                                = - \nabla \cdot \vec{g}~,
\end{equation}
\begin{equation}
\frac{\partial \vec{g}}{\partial t} =
      - \frac{B}{\rho} \nabla \delta\rho
      - \frac{K_A}{2}
           \vec{e}_z \times \nabla (\nabla^2 \vartheta_6)
      + \frac{\eta}{\rho} \nabla^2 \vec{g}
      + \frac{\zeta}{\rho} \nabla (\nabla \cdot \vec{g}) ~,
\end{equation}
\begin{equation}
\label{theta}
\frac{\partial \vartheta_6}{\partial t} =
                    \frac{1}{2\rho} \nabla \times \vec{g}
\end{equation}
\end{mathletters}

\noindent
as our final result for the hexatic phase. Apart from the absence of a
dissipative term in Eq.\ (\ref{theta}) it agrees with the findings of
Zippelius, Halperin and Nelson and therefore contains the same mode structure
as derived in that paper. For completeness sake, we mention however that the
equations of motion for the hexatic order parameter $\vartheta_6$ can be
derived from an effective action
\begin{equation}
S^{eff}[\vartheta_6] = \int_0^{\hbar \beta} d\tau \int d\vec{x}~
    \frac{1}{2} \vartheta_6
     \left( 4\rho \frac{\partial_{\tau}^2}{\partial^2}
       + 4i \eta \partial_{\tau} - K_A \partial^2
     \right) \vartheta_6~,
\end{equation}
that can easily be understood physically: The first term on the right-hand side
corresponds to the kinetic energy
$\int d\vec{x} \rho (\partial_{\tau} \vec{u})^2/2$
of the displacement field. The second term is a Leggett friction term and the
last term corresponds to the usual Frank energy, which is responsible for the
fact that the hexatic to liquid transition is of the Kosterlitz-Thouless type.

\section{HYDRODYNAMICS OF SUPERFLUID PHASES}
\label{SP}
Having arrived at the hydrodynamic equations for the solid and hexatic phases,
our next objective is to incorporate the effects of the additional hydrodynamic
degree of freedom associated with the phase of the superfluid order parameter.
Fortunately, from the microscopic theories developed for superfluid liquids
\cite{HM} and gases \cite{KD} it is well known how we should proceed to obtain
the hydrodynamic (two-fluid) equations for the superfluid phases starting from
the equations for the normal phase. The procedure consists in principle of four
steps. First, the total (average) density $\rho$ of the system is split up into
a normal density $\rho_n$ and a superfluid density $\rho_s$. In general these
densities are tensors of second rank, but for systems with hexagonal symmetry
which are of interest here they are proportional to the identity $\delta_{ij}$
and can be considered as scalars. Second, the total momentum density $\vec{g}$
is similarly split up into a normal component
$\rho_n \vec{v}_n$ and a superfluid component $\rho_s \vec{v}_s$ with a
superfluid velocity that is purely longitudinal
($\nabla \times \vec{v}_s = 0$). Third, for an effectively isotropic system the
dissipative terms in the momentum equation must be generalized to
\begin{eqnarray}
\eta \nabla^2 \vec{v}_n
      + \zeta_1 \frac{\rho_s}{\rho}
                \nabla (\nabla \cdot (\vec{v}_s - \vec{v}_n))
      + \zeta_2 \nabla (\nabla \cdot \vec{v}_n)~.    \nonumber
\end{eqnarray}
Finally, we must add the dynamics of the superfluid velocity, which is
basically determined from the Josephson relation and reads
\begin{equation}
\frac{\partial \vec{v}_s}{\partial t} =
    - \frac{B}{\rho^2} \nabla \delta\rho
    + \zeta_3 \frac{\rho_s}{\rho}
              \nabla (\nabla \cdot (\vec{v}_s - \vec{v}_n))
    + \zeta_4 \nabla (\nabla \cdot \vec{v}_n)~,
\end{equation}
where $B=\rho^2 d\mu/d\rho|_T$ is again the isothermal bulk modulus and
$\mu$ is the chemical potential per unit mass. We again leave out temperature
fluctuations since we are primarily interested in third-sound modes, for which
these fluctuations are (at least qualitatively) unimportant.

\subsection{Supersolid}
\label{HSS}
To apply the above procedure to Eq.\ (\ref{HS}) we must realize that we are
here in fact already dealing with a two-fluid hydrodynamics. We must therefore
not only split up the total momentum density $\vec{g}$ into a normal and a
superfluid component but also the net defect current, i.e.\
$\vec{J} = \vec{J}_n + \vec{J}_s$. Moreover, we have to account for the fact
that the chemical potential, just like the pressure, is a function of the
particle density and the net defect density. In this manner we arrive at the
following hydrodynamic equations
\begin{mathletters}
\begin{equation}
\frac{\partial \delta\rho}{\partial t} = -\nabla \cdot \vec{g}~,
\end{equation}
\begin{eqnarray}
\label{momrho}
\frac{\partial \vec{g}}{\partial t} =
      - \frac{B}{\rho} \nabla \delta\rho
     &-& \gamma_{\Delta} \nabla n_{\Delta}
      + 2\rho c_{\perp}^2 \nabla \times \vartheta_6
                                                   \nonumber \\
     &+& \eta \nabla^2 \vec{v}_n
      + \zeta_1 \frac{\rho_s}{\rho}
                \nabla (\nabla \cdot (\vec{v}_s - \vec{v}_n))
      + \zeta_2 \nabla (\nabla \cdot \vec{v}_n)~,
\end{eqnarray}
\begin{equation}
\frac{\partial \vartheta_6}{\partial t} =
                  \frac{1}{2\rho} \nabla \times \vec{g}~,
\end{equation}
\begin{equation}
\frac{\partial \vec{v}_s}{\partial t} =
    - \frac{B}{\rho^2} \nabla \delta\rho
    + \beta_{\Delta} \nabla n_{\Delta}
    + \zeta_3 \frac{\rho_s}{\rho}
              \nabla (\nabla \cdot (\vec{v}_s - \vec{v}_n))
    + \zeta_4 \nabla (\nabla \cdot \vec{v}_n)~,
\end{equation}
\begin{equation}
\frac{\partial n_{\Delta}}{\partial t} =
                  -\nabla \cdot \vec{J}~,
\end{equation}
\begin{equation}
\label{momdel}
\frac{\partial \vec{J}}{\partial t} =
      - c_{\Delta}^2 \nabla n_{\Delta}
      + \gamma \nabla \delta\rho
      + \kappa \nabla^2 \vec{J}_n
      + \xi_1 \nabla (\nabla \cdot \vec{J}_s)
      + \xi_2 \nabla (\nabla \cdot \vec{J}_n)~,
\end{equation}
\begin{equation}
\frac{\partial \vec{J}_s}{\partial t} =
    - \frac{B_{\Delta} \rho_s}{\rho^2} \nabla n_{\Delta}
    + \beta \rho_s \nabla \delta\rho
    + \xi_3 \frac{\rho_s}{\rho}
            \nabla (\nabla \cdot \vec{J}_s)
    + \xi_4 \nabla (\nabla \cdot \vec{J}_n)~,
\end{equation}
\end{mathletters}

\noindent
with
$\beta_{\Delta} = - \partial \mu/\partial
                           n_{\Delta}|_{\rho,T}$.
These represent nine equations for the nine unknown functions $\delta\rho$,
$\vec{v}_n$, $\vec{v}_s$, $\vartheta_6$, $n_{\Delta}$, $\vec{J}_n$ and
$\vec{J}_s$.

Although a complete analysis of the various hydrodynamic modes is now possible,
we will consider here only the situation which is most relevant to experiments,
namely that the normal part of the two-dimensional system is
clamped to an underlying substrate. As a result we have
$\vec{v}_n = \vec{J}_n = \vec{0}$ and
Eqs.\ (\ref{momrho}) and (\ref{momdel}) determining the normal properties of
the supersolid are no longer valid. The hydrodynamic equations therefore reduce
to
\begin{mathletters}
\label{clamped}
\begin{equation}
\frac{\partial^2 \delta\rho}{\partial t^2} =
    \frac{B \rho_s}{\rho^2} \nabla^2 \delta\rho
    - \beta_{\Delta} \rho_s \nabla^2 n_{\Delta}
    + \zeta_3 \frac{\rho_s}{\rho}
        \frac{\partial}{\partial t}(\nabla^2 \delta\rho)~,
\end{equation}
\begin{equation}
\frac{\partial^2 n_{\Delta}}{\partial t^2} =
    \frac{B_{\Delta} \rho_s}{\rho^2} \nabla^2 n_{\Delta}
    - \beta \rho_s \nabla^2 \delta\rho
    + \xi_3 \frac{\rho_s}{\rho}
        \frac{\partial}{\partial t} (\nabla^2 n_{\Delta})~.
\end{equation}
\end{mathletters}

\noindent
They contain two pairs of
propagating modes, which in the limit of a small coupling
constant
$\beta \ll BB_{\Delta}/\beta_{\Delta} \rho^4$ essentially correspond to
a pair of
third-sound modes with $\delta\rho$ unequal to zero but a constant net defect
density and a pair of modes with an oscillating net defect density.

One might have expected that the coupling of a superfluid
density to a propagating defect density  would have resulted
in one pair of gapped excitations and one pair of gapless excitations
instead.  Consider, for example, two identical superfluid layers.
If the layers are uncoupled the dynamics of the
phases $\vartheta_1$ and $\vartheta_2$ of the layers is determined by the
action
\begin{equation}
S_{layers}[\vartheta_1,\vartheta_2] =
 \int_0^{\hbar \beta} d\tau \int d\vec{x}~
   \left\{ \frac{\rho^2}{2B} (\partial_{\tau} \vartheta_1)^2 +
           \frac{\rho_s}{2} (\nabla \vartheta_1)^2
         + \frac{\rho^2}{2B} (\partial_{\tau} \vartheta_2)^2 +
           \frac{\rho_s}{2} (\nabla \vartheta_2)^2
   \right\}~,
\end{equation}
which clearly has two pairs of
gapless (third-sound) modes, one pair for each superfluid.
If we couple the order parameters by allowing the particles
to tunnel with an amplitude $-J/\rho$
from one layer to the other we must add
a Josephson coupling
\begin{equation}
S_{tunnel}[\vartheta_1,\vartheta_2] = - \int d\tau \int d\vec{x}~
    J \cos(\vartheta_1 - \vartheta_2)
\end{equation}
to this action.  The hydrodynamics modes couple to form two in-phase and
two out-of-phase excitations.  The
modes with $\vartheta_1$ and $\vartheta_2$ oscillating out of phase get
gapped (i.e.\
$\omega^2 \simeq BJ/\rho^2$ for $\vec{k}^2 \ll J/\rho_s$) and only the
modes with $\vartheta_1$ and $\vartheta_2$ oscillating in phase remain gapless.
Yet in Eq.\ (\ref{clamped}) we find only gapless modes.

This paradox can be resolved by noting that
we have made the standard assumption\cite{ZHN} that both the
total number of particles and the net number of defects is conserved. Hence,
after an atom has tunneled from a lattice site to the position of a vacancy, a
new vacancy is created near the original site of the atom.
The analogous process for the
two coupled superfluid layers in not simply tunneling of
individual atoms from one layer to another, but rather the
exchange of a pair of atoms in different layers, returning the
system to its original state. Such a process is not a
Josephson coupling and therefore the modes remain gapless.
The existence of separate conservation laws for the particle and defect
density  thus allows in principle two separate broken symmetries.

We also note in passing
that the third-sound modes in Eq.\ (\ref{clamped}) are not present
in the hydrodynamic equations proposed by Andreev and Lifshitz \cite{A} and
considered in more detail by Liu. \cite{L} This is a result of the fact that
these authors use a somewhat different physical picture for the supersolid
phase: They assume that the superfluid current density is carried by (Bose
condensed) defects and that the normal current density is solely due to lattice
vibrations. Hence if we take
$\vec{v}_n = \vec{0}$, which in their context means that $\partial
\vec{u}/\partial t = \vec{0}$, only transport of defects is possible and only
the latter two modes survive. However, as a consequence of their picture the
hydrodynamic equations in the (normal) solid phase describe only longitudinal
and transverse sound modes in an ideal lattice and do not include the effect of
vacancies or interstitials. As explained above this is incorrect in principle
and one should at least also allow for a normal current density due to the
motion of defects. In addition, we have seen in Sec.\ \ref{SOP} that even in
the presence of defects the density fluctuations are equal to $-\rho
u^{Phys}_{ii}$. It is therefore perfectly reasonable that if there is
superfluid mass transport possible in the solid, it can be caused both by the
motion of defects and by lattice vibrations. Indeed, as an existence proof of
this latter possibility we can for instance consider superfluid $^4$He in a
weak periodic and commensurate potential, which is clearly a supersolid without
defects.

While it is generically possible to have both density and
defect superfluid modes, we might expect however, for realistic films on
realistic substrates, that in a supersolid it may be harder for particles to
perform ring exchanges\cite{K2,ring} than for vacancies
to exchange positions. Thus, {\it a priori}, we might expect the effective
superfluid stiffness for the density fluctuations to be smaller than that
of the vacancies, perhaps to the point where the former is entirely absent.

\subsection{Superhexatic}
We next turn to the superhexatic phase. In a similar manner as in Sec.\
\ref{HSS} we obtain from Eq.\ (\ref{HH}) the full set of hydrodynamic equations
\begin{mathletters}
\begin{equation}
\frac{\partial \delta\rho}{\partial t}
                                = - \nabla \cdot \vec{g}~,
\end{equation}
\begin{eqnarray}
\frac{\partial \vec{g}}{\partial t} =
      - \frac{B}{\rho} \nabla \delta\rho
     &-& \frac{K_A}{2}
           \vec{e}_z \times \nabla (\nabla^2 \vartheta_6)
                                                  \nonumber \\
     &+& \eta \nabla^2 \vec{v}_n
      + \zeta_1 \frac{\rho_s}{\rho}
                \nabla (\nabla \cdot (\vec{v}_s - \vec{v}_n))
      + \zeta_2 \nabla (\nabla \cdot \vec{v}_n)~,
\end{eqnarray}
\begin{equation}
\label{theta6}
\frac{\partial \vartheta_6}{\partial t} =
                    \frac{1}{2\rho} \nabla \times \vec{g}~,
\end{equation}
\begin{equation}
\frac{\partial \vec{v}_s}{\partial t} =
    - \frac{B}{\rho^2} \nabla \delta\rho
    + \zeta_3 \frac{\rho_s}{\rho}
              \nabla (\nabla \cdot (\vec{v}_s - \vec{v}_n))
    + \zeta_4 \nabla (\nabla \cdot \vec{v}_n)~,
\end{equation}
\end{mathletters}

\noindent
that leads to the usual two-fluid hydrodynamics of a superfluid if we omit Eq.\
(\ref{theta6}) and put $\vartheta_6 = 0$. Therefore these equations allow for
first and second sound, \cite{second} and for a pair of transverse modes
involving $\vec{v}_n^T$ and $\vartheta_6$ which are either dispersive or
propagating depending on the sign of
$\Delta = K_A/4\rho - (\eta/\rho_n)^2$: If $\Delta \leq 0$ we have two
purely dispersive modes with
$\omega^{\pm} = -i(\eta/\rho_n \pm \sqrt{-\Delta}) \vec{k}^2/2$,
wheras if $\Delta > 0$ we have two propagating modes and the particle-like
dispersion
$\omega^{\pm} = \pm \sqrt{\Delta}~ \vec{k}^2/2
                            - i (\eta/\rho_n) \vec{k}^2/2$.
However, considering again the case $\vec{v}_n = \vec{0}$ the hydrodynamic
equations now simply reduce to
\begin{equation}
\frac{\partial^2 \delta\rho}{\partial t^2} =
    \frac{B \rho_s}{\rho^2} \nabla^2 \delta\rho
    + \zeta_3 \frac{\rho_s}{\rho}
        \frac{\partial}{\partial t}(\nabla^2 \delta\rho)~,
\end{equation}
which contains only a pair of third-sound modes with the velocity
$c_3 = \sqrt{B\rho_s/\rho^2}$ and the diffusion constant
$D_3 = \zeta_3 \rho_s/\rho$.

\section{CONCLUSIONS AND DISCUSSION}
\label{DC}
In this paper we have derived the hydrodynamic equations for the  supersolid
and superhexatic phases of a neutral two-dimensional Bose fluid. For the
supersolid these equations are rather complex, since they incorporate the
effects of defect motion and lattice vibrations on both the normal and
superfluid parts of the momentum density. Our physical picture for the
influence on the superfluid part is roughly speaking that in a mean-field
theory the condensate wavefunction $\Psi(\vec{x},t)$ obeys the Schr\"odinger
equation
\begin{equation}
i\hbar \frac{\partial \Psi(\vec{x},t)}{\partial t} =
   \left\{
     - \frac{\hbar^2 \nabla^2}{2m}
     + \int d\vec{x}'~ V(\vec{x}-\vec{x}') n(\vec{x}',t)
   \right\} \Psi(\vec{x},t)~,
\end{equation}
where $m$ is the mass of the Bose particles and $V(\vec{x}-\vec{x}')$ is their
interaction. In addition, $n(\vec{x},t)$ is the particle density which will be
determined by an additional mean-field theory that, for a supersolid, shows the
instability associated with the formation of a density wave. Hence the
(thermal) average
$\langle n(\vec{x},t) \rangle$ is periodic in space and independent of time. As
a result the condensate wavefunction is, if we neglect density fluctuations,
also periodic and we have indeed both diagonal as well as off-diagonal
long-range order. Fluctuations in the density, however, induce variations in
the phase of the wavefunction and therefore in the superfluid velocity. Because
these density fluctuations can be caused by both lattice vibrations and
oscillations in the net defect density we conclude that both mechanisms can
lead to superfluid motion. Together with the existence of a conservation law
for the net number of defects, this explains from a more microscopic view
why we found two third-sound modes and two modes with an oscillatory net defect
density in the case of a supersolid adsorbed onto a substrate.

For the superhexatic phase we have shown that the hexatic long-range order
leads to an additional (as compared to the superfluid) hydrodynamic degree of
freedom that affects only the transverse modes and is therefore at long
wavelengths decoupled from the superfluid momentum density. This can also be
understood from the above picture, since variations in the orientational order
parameter $\vartheta_6$ do not lead to density fluctuations in first instance.
As a result we find on a substrate only two third-sound modes and thus at the
hydrodynamic level of description nothing to distinguish the superhexatic from
the superfluid. Although this is in agreement with the experiments of Chen and
Mochel, who indeed only observe one third-sound branch below the second
critical temperature $T_c$, it is unfortunate for the purpose of suggesting a
possible identification of the superhexatic phase. On the basis of our results
we can, however, conclude that a more microscopic probe is needed if one wants
to detect the orientational order present in a superhexatic helium film. In our
opinion this appears to be an important, but also difficult experimental
challenge.

Finally, we would also like to point out the possible relevance of our results
to the recent experiments with bulk solid $^4$He. \cite{LG} In these
experiments Lengua and Goodkind observe at sufficiently high frequencies an
additional (resonant) attenuation and velocity change of sound. Moreover, they
notice that their data can be explained by a simple model of two coupled wave
equations which turns out to be identical to the longitudonal part of our
solid hydrodynamics derived in
Sec.\ \ref{SOP}. Because our two-dimensional hydrodynamics should be able to
describe the propagation of sound perpendicular to the c-axis of hcp $^4$He,
this confirms the conjecture of Lengua and Goodkind that the collective mode
observed is associated with the motion of defects. For a more detailed
discussion of the coupling between sound and the defects one should of course
consider the fully three-dimensional situation and include the anisotropy of
the hcp crystal. Work in this direction is in progress.

\section*{ACKNOWLEDGMENTS}
This research was supported by Grant No.\ DMR-9502555 and DMR-9416906
from the National
Science Foundation, the ESF Network on Quantum Fluids and Solids,
the Swedish Natural Science Research Council and the
Stichting voor Fundamenteel Onderzoek der Materie (FOM) which is
financially supported by the Nederlandse Organisatie voor
Wetenschappelijk Onderzoek (NWO).
We thank Huug van Beelen, Henk van Beijeren, Michel Bijlsma,
Reyer Jochemsen, and Anne van Otterlo for stimulating and helpful discussions.

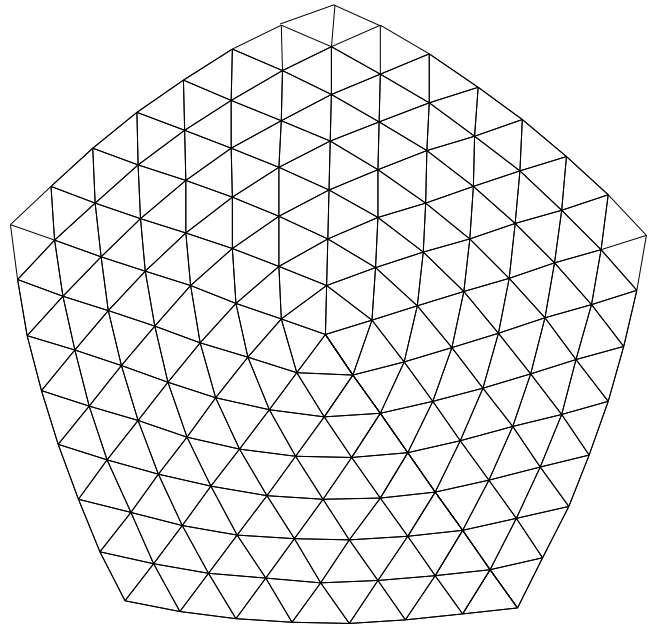
\begin{figure}
\caption{A negative disclination in a hexagonal lattice. It is
         formed by removing a $\rm 60^\circ$ wedge from the
         lattice, and then distorting the lattice so that the
         open edges meet. A positive disclination (not shown)
         is formed by the insertion of a $\rm 60^\circ$ wedge
         of material into the lattice.
         \label{fig1}}
\end{figure}
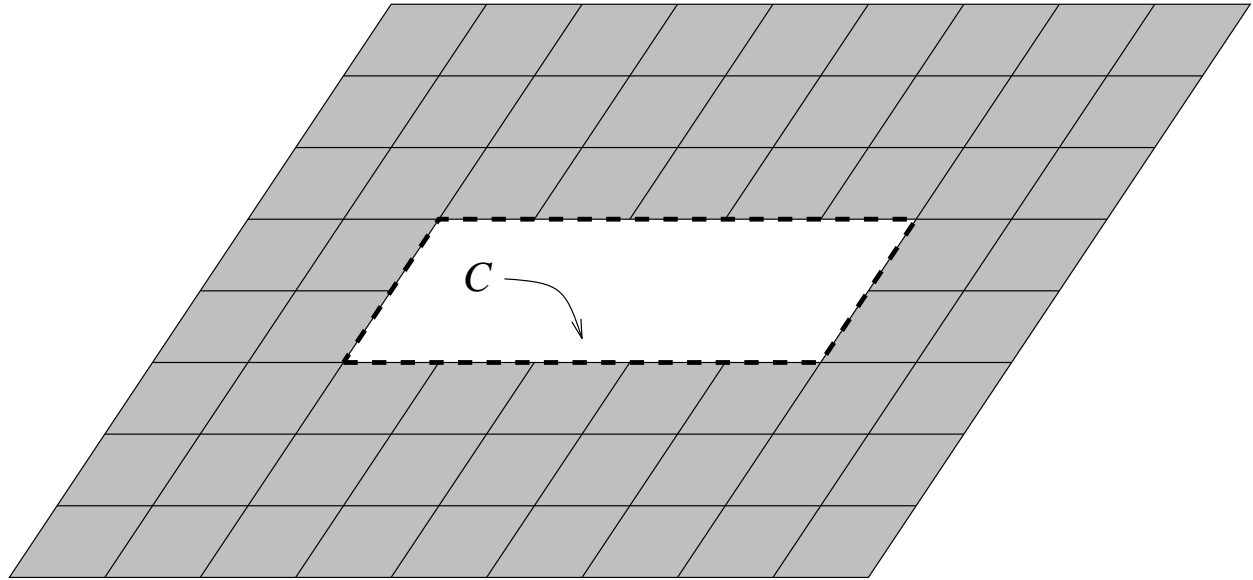
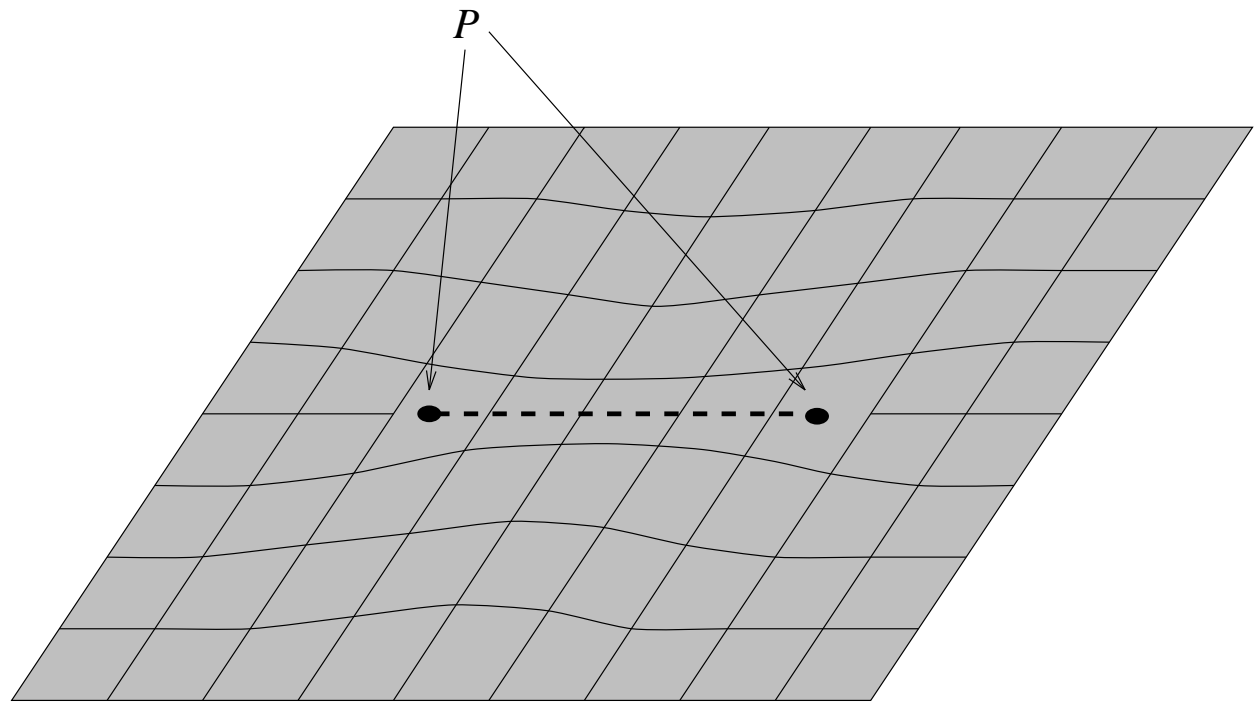
\begin{figure}
\caption{A Volterra construction for a pair of dislocations:
         (a) a section of the two-dimensional lattice
             bounded by the loop $\cal C$ is removed and
         (b) the edges of the loop are pulled together to form
             the dashed line $\cal L$. The endpoints $\cal P$
             of this line correspond to the positions of the
             dislocations.
         \label{fig2}}
\end{figure}
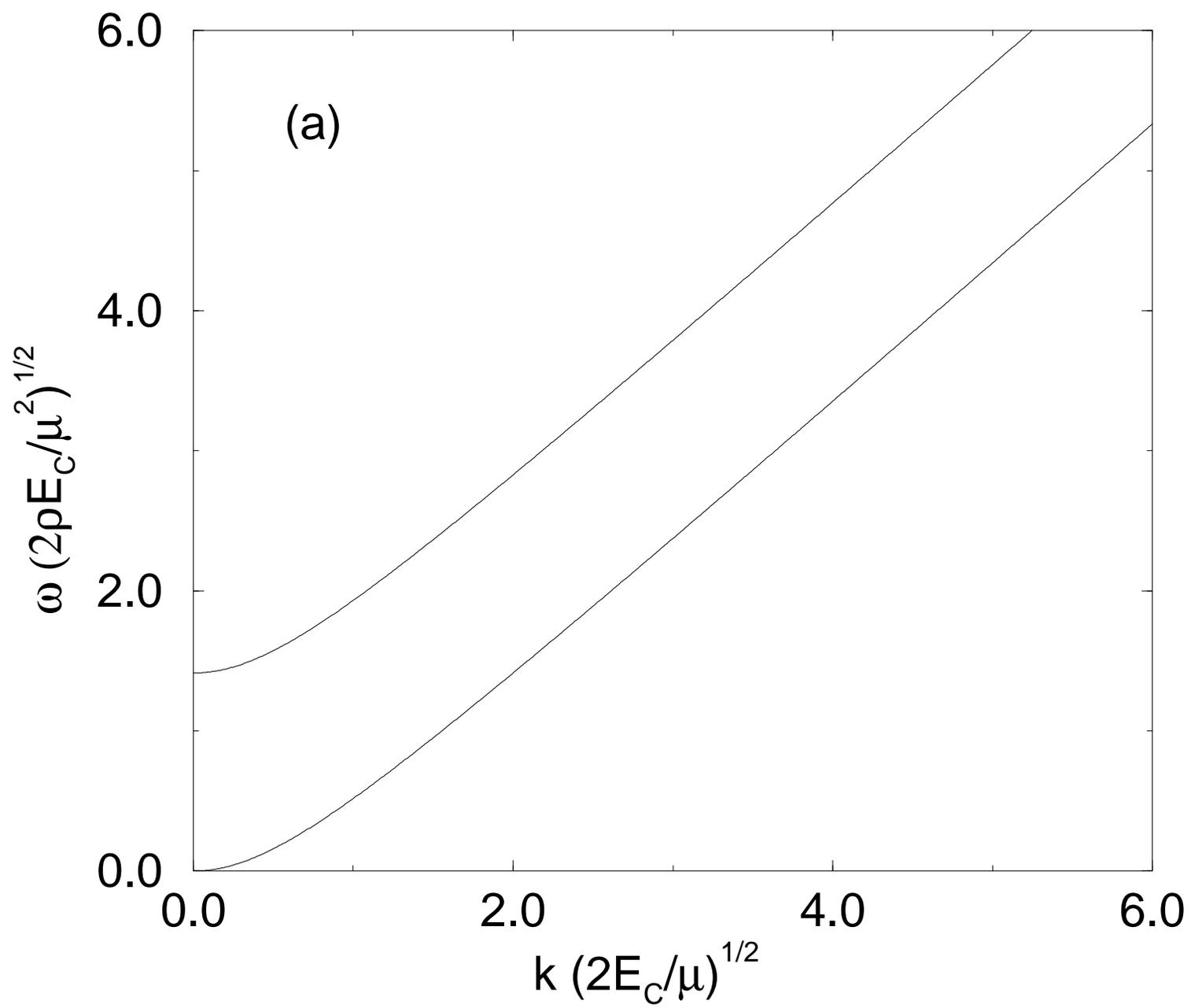
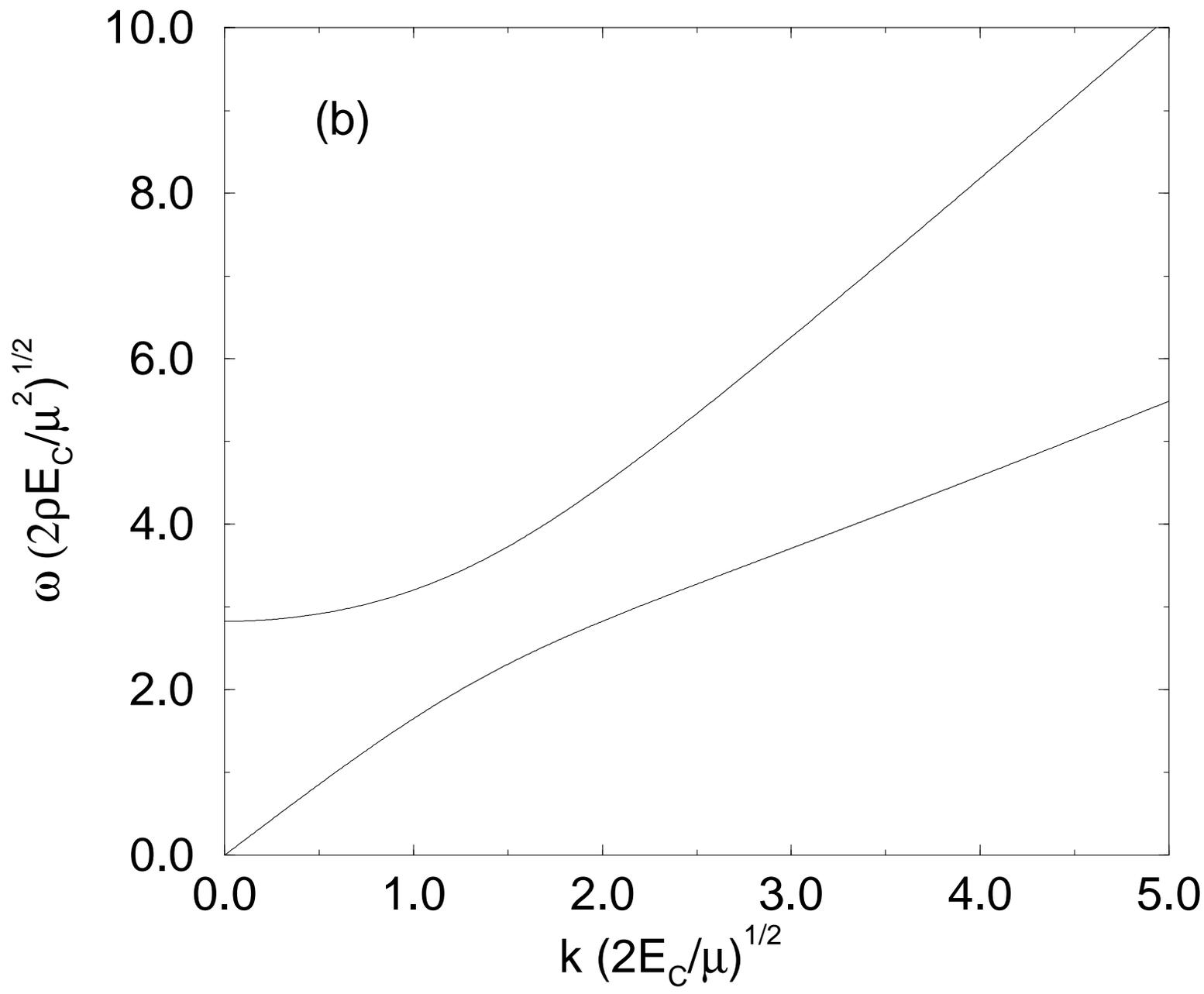
\begin{figure}
\caption{Dispersion curves for (a) the transverse and (b) the
         longitudinal modes in the hexatic phase.
         In both cases $\nu = 1/2$.
         \label{fig3}}
\end{figure}
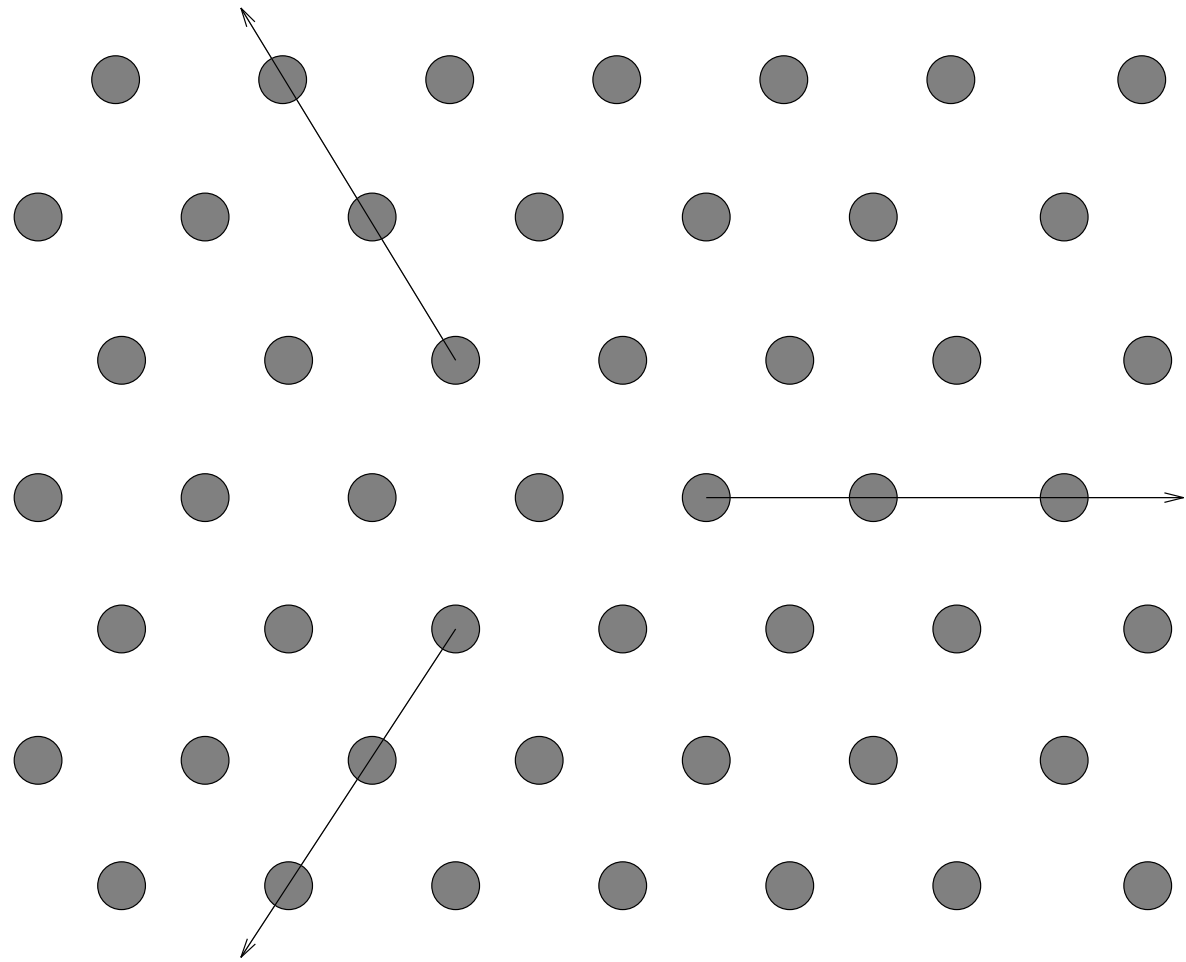
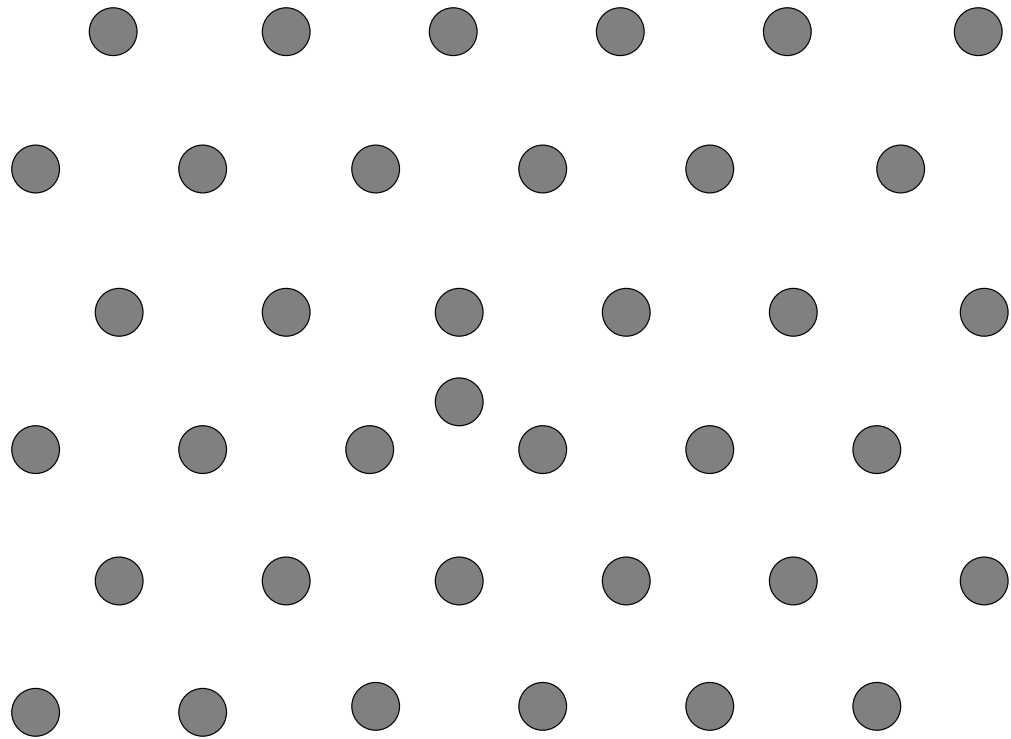
\begin{figure}
\caption{A Volterra construction for an interstitial in a
         hexagonal lattice:
         (a) three half-infinite lines of lattice points
             are removed (as indicated by the arrows) and
         (b) the lattice is distorted by drawing
             together the edges about the removed lines in such
             a manner that the crystal symmetry at large
             distances from the interstitial is restored.
             The removal of the three half-infinite lines is
             simply the Volterra construction of three
             dislocations. Thus an interstitial can be viewed as
             being made from three dislocations.
             A similar construction can be made for vacancies.
         \label{fig4}}
\end{figure}


\begin{references}
\bibitem{bulk} P. Kapitza, Nature {\bf 141}, 74 (1938); and
               J.F. Allen and A.D. Misener,
               Nature {\bf 141}, 75 (1938).
\bibitem{film} I. Rudnick,
               Phys. Rev. Lett. {\bf 40}, 1454 (1978); and
               D.J. Bishop and J.D. Reppy,
               Phys. Rev. Lett {\bf 40}, 1727 (1978).
\bibitem{Mochel} M.T. Chen, J. Roesler, and J.M. Mochel,
                 J. Low Temp. Phys. {\bf 89}, 125 (1992); and
                 J.M. Mochel and M.T. Chen,
                 Phys. B {\bf 197}, 278 (1994).
\bibitem{VAV} For a different interpretation of these experiments
              see S.-C. Zhang, Phys. Rev. Lett. {\bf 27}, 2142
              (1993); and M. Gabay and A. Kapitulnik,
              Phys. Rev. Lett. {\bf 71}, 2138 (1993).
\bibitem{NK} D.R. Nelson and J.M. Kosterlitz,
             Phys. Rev. Lett. {\bf 39}, 1201 (1977).
\bibitem{PRL} K. Mullen, H.T.C. Stoof, M. Wallin, S.M. Girvin,
              Phys. Rev. Lett. {\bf 72}, 4013 (1994).
\bibitem{NH} D.R. Nelson and B.I. Halperin,
             Phys. Rev. B {\bf 19}, 2457 (1979).
\bibitem{Anne} A. van Otterlo and K.-H. Wagenblast,
               Phys. Rev. Lett. {\bf 72}, 3598 (1994)
               and references therein.
\bibitem{LG} G.A. Lengua and J.M. Goodkind,
             J. Low. Temp. Phys. {\bf 79}, 251 (1990).
\bibitem{B} L. Balents (unpublished).
\bibitem{Z} S.C. Zhang, Int. J. Mod. Phys. B {\bf 6}, 25 (1992).
\bibitem{fluid} In this paper we will at times use the adjective
                `superfluid' for those properties that
                superfluids, superhexatics and supersolids have
                in common, although we realize that hexatic
                and solid phases are not fluid.
\bibitem{K1} H. Kleinert, J. Phys. A {\bf 19}, 1855 (1986).
\bibitem{La} L.D. Landau and E.M. Lifshitz, {\it Theory
             of Elasticity} (Pergamon, New York, 1970).
\bibitem{K2} H. Kleinert, {\it Gauge Fields in Condensed Matter
             Physics} (World Scientific, Singapore, 1989).
\bibitem{Na} F.R.N. Nabarro, {\it Theory of Dislocations}
             (Clarendon, New York, 1967), Chapter VII.
\bibitem{K3} H. Kleinert, Phys. Lett. A {\bf 91}, 295 (1982).
\bibitem{M} P.C. Martin, O. Parodi, and P.S. Pershan,
            Phys. Rev. A {\bf 6}, 2401 (1972).
\bibitem{ZHN} A. Zippelius, B.I. Halperin, and D.R. Nelson,
              Phys. Rev. B {\bf 22}, 2514 (1980).
\bibitem{N} D.R. Nelson (unpublished).
\bibitem{TL} A.O. Caldeira and A.J. Leggett,
             Ann. Phys. {\bf 149}, 374 (1983); and
             A.J. Leggett, S. Chakravarty, A.T. Dorsey,
             M.P.A. Fisher, A. Garg, and W. Zwerger,
             Rev. Mod. Phys. {\bf 59}, 1 (1987).
\bibitem{HM} P.C. Hohenberg and P.C. Martin,
             Ann. Phys. {\bf 34}, 291 (1965).
\bibitem{KD} T.R. Kirkpatrick and J.R. Dorfman,
             J. Low. Temp. Phys. {\bf 58}, 301 (1985);
             {\it ibid.} {\bf 58}, 399 (1985).
\bibitem{A} A.F. Andreev and L.M. Lifshitz,
            Sov. Phys. JETP {\bf 29}, 1107 (1969).
\bibitem{L} M. Liu, Phys. Rev. B {\bf 18}, 1165 (1978).
\bibitem{ring} R.P. Feynman, Phys. Rev. {\bf 91}, 1291 (1953).
\bibitem{second} For a complete description of second sound one
                 should of course also include temperature
                 fluctuations.
\end{references}
\end{document}